\documentclass[a4paper,11pt,article,oneside]{memoir}
\usepackage{ecis2019}

\usepackage[left=2cm, right=2cm, lines=45, top=0.8in, bottom=1in]{geometry}

\usepackage{graphicx} 
\usepackage{csquotes}
\usepackage{enumitem}
\usepackage[center]{caption}
\usepackage[section]{placeins}
\usepackage{multirow}

\maintitle{Convolutional Neural Networks with A Topographic Representation Module for EEG-Based Brain-Computer Interfaces}

\authors{
Xinbin Liang, National University of Defense Technology, Changsha, China, lxb2203@163.com

Yaru Liu, National University of Defense Technology, Changsha, China, lyrnudt@163.com

Yang Yu, National University of Defense Technology, Changsha, China, yuyangnudt@hotmail.com

Kaixuan Liu, National University of Defense Technology, Changsha, China, 2314825269@qq.com

Yadong Liu, National University of Defense Technology, Changsha, China, liuyadong1977@163.com

Zongtan Zhou, National University of Defense Technology, Changsha, China, narcz@163.com
}

\shortauthors{Xinbin Liang, et al.} 

\usepackage{amsmath}

\begin{document}
\vspace{-1.5cm}
\begin{center}
\end{center}
\vspace{1cm}
\begin{abstract}\noindent
Objective: Convolutional Neural Networks (CNNs) have shown great potential in the field of Brain-Computer Interfaces (BCIs). The raw Electroencephalogram (EEG) signal is usually represented as 2-Dimensional (2-D) matrix composed of channels and time points, which ignores the spatial topological information. Our goal is to make the CNN with the raw EEG signal as input have the ability to learn EEG spatial topological features, and improve its performance while essentially maintaining its original structure. 
Methods:We propose an EEG Topographic Representation Module (TRM). This module consists of (1) a mapping block from the raw EEG signal to a 3-D topographic map and (2) a convolution block from the topographic map to an output of the same size as input. According to the size of the kernel used in the convolution block, we design 2 types of TRMs, namely TRM-(5,5) and TRM-(3,3). We embed the TRM into 3 widely used CNNs, and tested them on 2 publicly available datasets (Emergency Braking During Simulated Driving Dataset (EBDSDD), and High Gamma Dataset (HGD)). 
Results: The results show that the classification accuracies of all 3 CNNs are improved on both datasets after using the TRM. With TRM-(5,5), the average  accuracies of DeepConvNet, EEGNet and ShallowConvNet are improved by 6.54\%, 1.72\% and 2.07\% on EBDSDD, and by 6.05\%, 3.02\% and 5.14\% on HGD, respectively; with TRM-(3,3), they are improved by 7.76\%, 1.71\% and 2.17\% on EBDSDD, and by 7.61\%, 5.06\% and 6.28\% on HGD, respectively.
Significance: We improve the classification performance of 3 CNNs on 2 datasets by the use of TRM, indicating that it has the capability to mine the EEG spatial topological information. In addition, since the output of TRM has the same size as the input, CNNs with the raw EEG signal as input can use this module without changing their original structures.
\end{abstract}

\begin{keywords}
 Convolutional Neural Network (CNN), Electroencephalogram (EEG), topographic representation, Brain-Computer Interface (BCI), EEG decoding, deep learning, ShallowConvNet, DeepConvNet, EEGNet.
\end{keywords}

\chapter{Introduction}

BCIs enable direct communication between human and machine via EEG (\cite{1}). EEG signal contains instinctive biometric information from the human brain. Through the precise EEG decoding, BCIs can recognize the user’s inner thoughts. In general, the EEG decoding consists of 5 main stages: data collection, signal pre-processing, feature extraction, classification and data analysis (\cite{2}). Although these stages are essentially the same in a BCI paradigm, signal pre-processing (\cite{3}), feature extraction (\cite{4}), and classification methods (\cite{5}) typically require substantial expertise and some prior knowledge about the BCI paradigm. Moreover, due to the manual processing, some useful information may be excluded from the extracted features, which poses a challenge to the subsequent classification and data analysis.

Deep learning has largely alleviated the need for manual feature extraction with the combination of feature extraction and classification. CNNs, in particular, have achieved great success in many challenging image classification tasks, outperforming approaches that rely on hand-crafted features (\cite{6,7}). Inspired by the success of deep learning in areas such as computer vision and natural language processing, researchers have introduced it to the EEG decoding area (\cite{8}). CNNs are one of the most versatile deep learning methods in the BCI. Among all deep learning-based EEG decoding methods, those related to CNNs accounted for 53\% of the total (43\% for CNNs and 10\% for hybrid-CNNs) (\cite{9}). CNNs are typically composed of 3 structure blocks: convolutional layers, pooling layers and fully connected layers. The convolutional layer is an essential part of the CNN, which performs the feature extraction function. The pooling layer provides a down-sampling operation that both ensures learning of more robust features and reduces computation. The fully connected layer is typically located at the bottom of the network and implements the combined local features and classifier functions. The architecture of a CNN generally consists of layers arranged in a specific order, with earlier layers learning lower-level features and deeper layers learn high-level features. Several studies have use CNN models,  including light (\cite{10,11,12,13,14,15}) and deep (\cite{12,16,18,19}) architectures, as well as other varieties (\cite{20,21,22,23,24,25,26}), to decode EEG signals. Waytowich et al. introduced a compact CNN for feature extraction and classification from the raw Steady-State Visually Evoked Potential (SSVEP) signal directly, with an average accuracy of about 80\% cross-subject on a 12-classs dataset (\cite{14}). By introducing batch normalization in the input and convolutional layers to cope with the overfitting problem, Liu et al. applied a CNN to the detection task of P300 signals, and achieved state-of-the-art recognition performance on both Dataset IIb of BCI competition II and Dataset II of BCI competition III (\cite{15}). Tang et al. proposed a CNN model based on spatial-temporal features to classify single Motor Imagery (MI) tasks, and the results showed that compared with traditional methods, CNN can further improve classification performance (\cite{19}). To address the overfitting problem of traditional machine learning methods in EEG-based emotion reading, Li et al. used a hierarchical CNN with differential entropy features from different channels as input, and achieved classification results with advantages compared to traditional methods (\cite{26}). Li et al. proposed MI-VGG by modifying the VGG network to enable effective recognition of spectral images generated by MI-EEG and obtained competitive results on 3 publicly available datasets (\cite{16}).

EEG-based CNNs use both raw signals and features generated from raw signals as input. In this paper, we focus only on CNNs with the raw 2-D EEG signal as input. The raw EEG signal refers to the EEG data in the time domain, i.e., the [C (Channels) × TP (Time Points)] matrix. Since deep learning based CNN models have the ability to learn complex features from data without using hand-crafted features and can achieve end-to-end learning, the raw EEG signal is the most commonly used input formulation (\cite{27}). Based on the classification error, CNNs learn and optimize the feature representation of the raw EEG signal simultaneously. Several competitive CNN models using raw EEG signals as input have been proposed (\cite{11,12,19,24,28,29,30,31,32,33}). Schirrmeister et al. investigated more systematically the end-to-end learning from raw signals in EEG decoding using CNNs, and designed 2 widely used network architectures, DeepConvNet and ShallowConvNet (\cite{12}). Test results on 2 different datasets showed that the proposed networks achieved at least as good classification performance as the best traditional methods. In addition, the visualization of the learned features also showed that the 2 networks performed an effective spatial mapping. Lawhern et al. introduced a compact CNN, EEGNet, which uses depth-wise and separable convolutions. Test results on 4 different types of BCI paradigms showed that EEGNet had better generalization ability, while obtaining comparable classification performance to others methods (\cite{11}). Amin et al. used the raw EEG signal without preprocessing or artifact removal as input, and the classification performance was significantly improved by fusing multiple CNN models with different architectures (\cite{32}). CNNs with the raw EEG signal as input, ignore the spatial inter-topology of the electrodes; therefore, most of these networks contain a spatial (depth) convolutional layer to learn the weights of the electrodes, which is equivalent to a compensatory operation for ignoring the EEG spatial topological information.

In this paper, we introduce a Topographic Representation Module (TRM) to address the EEG spatial topological information loss problem caused by CNNs using raw EEG signals as input. The TRM consists of (1) a mapping block form the raw EEG signal to the topographic map and (2) a convolutional block transforming the topographic map into an output the same size as the input. That is, the size of the EEG signal will remain unchanged after passing through our TRM. Any CNN that takes the raw 2-D EEG signal as input can use this module. Such a design makes use of both the EEG spatial topological information and various existing excellent CNN architectures, making the TRM very versatile.
The rest of this paper is organized as follows. Section 2 presents materials and methods, including datasets, classification algorithms, the TRM, implementation details and evaluation metrics. Section 3 presents the experimental results, i.e., algorithms performance under different evaluation metrics. Section 4 presents a discussion of the algorithms and the results. Finally, we give a conclusion.

\chapter{MATERIALS AND METHODS}

\section{Datasets}
The EEG signals in BCIs are generally classified into two types according to the presence or absence of external stimuli: evoked potentials and spontaneous EEG (\cite{2}). Evoked-potential BCIs have a clear external stimulus, and the EEG signal exhibits certain time-locked characteristics. Usually, this type of BCI has a high classification accuracy. Unlike evoked-potential BCIs, spontaneous EEG-based BCIs rely on the subject’s spontaneous brain activity, usually without an external stimulus, and are generally more difficult to train. In this paper, we choose an evoked-potential BCI dataset and a spontaneous EEG-based BCI dataset, respectively.

\textbf{Dataset 1: Emergency Braking During Simulated Driving Dataset (EBDSDD).}
 EBDSDD has been described in detail in (\cite{34}). Here, we briefly describe it as follows. EBDSDD is a dataset of 2 types of tasks (emergency braking and non-braking) obtained from 18 subjects, each performing approximately 210 emergency braking (target) trials. EEG signals are recorded using 59 electrodes placed at the scalp sites, low-pass filtered at 45 Hz, and down-sampled to a sampling rate of 200 Hz. We refer to (\cite{34}) for the data processing method. The data from 1300 ms before emergency braking to 200 ms after braking are chosen as target segments. Non-target segments are normal driving EEG signals of length 1500 ms away from any stimulation and braking behavior for at least 3000 ms. Baseline correction is performed segment-wise using the data from the first 100 ms. For each subject, we choose the same number of the non-target segments and the target segments. In addition, Electrodes FP1, FP2, AF3 and AF4 are susceptible to the interference from the oculomotor potential, which we exclud and use only the remaining 55 electrodes. Therefore, the size of each target and non-target segment is 55 (channels) × 280 (time points). 

\textbf{Dataset 2: High Gamma Dataset (HGD).}
 HGD has been described in detail in (\cite{12}). Here, we give a brief description. HGD is a dataset of 4-class movements (left hand, right hand, feet and rest) obtained from 14 healthy subjects, each with approximately 1000 4-second trials of executed movements. The first approximately 880 trials are the training set and the last approximately 160 trials are the test set. HGD is a 128-electrode dataset with a sampling rate of 500 Hz.For the data processing method, we refer to (\cite{12}). 44 electrodes covering the motor cortex (all central electrodes except Cz, which is used as the recording reference electrode) are selected. The EEG signal is filtered using a 4-125 Hz bandpass filter and down-sampled to 250 Hz. We adopt the standard trial-wise training strategy, using the entire duration of the trial, so the matrix for each trial data is 44 (channels) × 1000 (time points). 

\section{Classification Algorithms}
In this paper, we use 3 widely used representative EEG-based CNNs: ShallowConvNet, DeepConvNet and EEGNet. They are CNNs specially designed for EEG decoding, and have shown better performance than traditional methods in many BCI applications.

\textbf{ShallowConvNet.}
The design of ShallowConvNet is inspired by the Common Spatial Patterns of Filter Banks (FBCSP). It is similar to FBCSP in terms of EEG feature extraction. ShallowConvNet has a simple architecture, only consisting of a temporal convolution layer, a spatial convolution layer, an average pooling layer and a dense layer. It has achieved better results than the best traditional methods in many EEG decoding tasks (\cite{11,12,18,35,36,37,38,39}). For more details about ShallowConvNet, please refer to (\cite{12}).

\textbf{DeepConvNet.}
The architcture of DeepConvNet is inspired by the successful architecture of deep CNNs in computer vision, which aims to extract a wide range of features without relying on specific features types. DeepConvNet is designed as a general CNN architecture with the hope of achieving competitive accuracy with only a small amount of expert knowledge. It consists of four Conv-Pool blocks and a dense classification layer. The first Conv-Pool block is divided into a temporal convolutional layer, a spatial convolutional layer and a max-pooling layer. The other Conv-Pool blocks consist of only one convolutional layer and one max-pooling layer. DeepConvNet has achieved competitive classification accuracy compared to traditional methods in many EEG decoding tasks (\cite{11, 12, 18, 35, 37, 39, 40}). For more details about DeepConvNet, please refer to (\cite{12}).

\textbf{EEGNet.}
EEGNet is designed to find a single CNN architecture that could be applied to different types of EEG-based BCIs, and make the network as compact as possible. The structure of EEGNet consists of 2 Conv-Pool blocks and a classification layer. In Block 1, it performs a temporal convolution and a depth-wise convolution sequentially. In Block 2, it uses separable convolution consisting of a depth-wise convolution and a point-wise convolution. In the classification layer, the SoftMax method is used. Due to the use of depth-wise and point-wise convolutions and the omission of the dense layer, this network design reduces the trainable parameters of EEGNet by at least an order of magnitude compared to other CNNs. Related studies have shown that EEGNet has a reasonable structure and excellent performance in different types of BCI paradigms (\cite{11,18,35,36,37,38,39,40,41,42,43,44,45}). For a detailed description of EEGNet, please refer to (\cite{11}).

\section{Topographic Representation Module}
\begin{figure}[t]
\begin{center}
  \includegraphics[width=\textwidth]{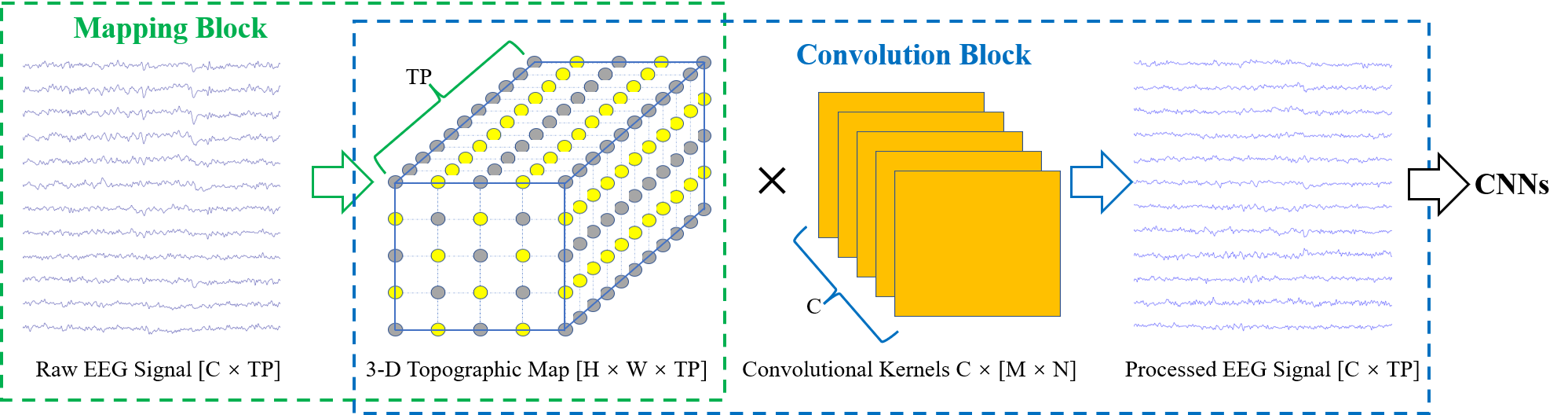}
  \caption{Overall visualization of the EEG Topographic Representation Module (TRM). C: Channels, TP: Time Points, H: Height, W: Width, [M×N]: Kernel size. The parts framed by green and blue dashed lines are the mapping block and the convolution block, respectively. In the 3-D Topographic Map, the coordinates of the yellow points correspond to the electrode locations of the raw EEG signal, and the values are the potential values of the electrodes, while the gray points have no corresponding electrodes and the values are set to 0. The output and input of the TRM are of the same size.}
  \label{fig:Fig_1}
\end{center}
\end{figure}

In order to make the CNN with the raw EEG signal as input more effective in utilizing the electrode spatial topological information, and to take advantage of various excellent EEG-based CNNs, we propose the EEG Topographic Representation Module (TRM). This module consists of (1) a mapping block from the raw EEG signal to the 3-D topographic map and (2) a convolution block from the 3-D topographic map to the processed output with the same size as the input, as shown in Figure \ref{fig:Fig_1}. 

\textbf{Mapping Block.}
According to the correspondence between the channels and electrode locations on the scalp, the raw EEG signal with the size of [C × TP] is mapped into a 3-D EEG topographic map with the size of [H (Height) × W (Width) × TP (Time Points)]. For the correspondence between the electrodes and the 2-D matrix coordinates of size [H × W], we refer to (\cite{46,47}), and adjust the size of the matrix according to the distribution of electrodes. After mapping, the potential value of the raw EEG signal remains unchanged on the 2-D matrix, and we set them to 0 for coordinates without corresponding electrodes on the matrix. In this way, we transform the raw EEG signal at a certain time into a 2-D topographic map represented by electrode locations (matrix coordinates) and their corresponding potentials. These 2-D topographic maps are combined in temporal order to form a 3-D topographic map.

\textbf{Convolution Block.}
We use the convolution on the 3-D EEG topographic map to take advantage of its powerful feature learning capability. In addition, to take advantage of the various excellent architectures of EEG-based CNNs, we make the output of the TRM the same size as the input so that it can be embedded into these networks. According to the size of the convolutional kernels used in the convolutional blocks, we design two types of TRMs, namely TRM-(5, 5) and TRM-(3, 3). In TRM-(5, 5), the first layer uses C convolution kernels of size [5 × 5] with a step of 1. The size of the convolution kernels in the subsequent layers is chosen according to the size of the feature map after convolution in the previous layer. If the feature map is larger than [5 × 5], C convolution kernels of size [5 × 5] are continued to be used, otherwise, C convolution kernels of the same size as the feature map are used. Similarly, in TRM-(3, 3), the first layer uses C convolution kernels of size [3 × 3] with a step size of 1. The size of the convolution kernels in the subsequent layers is chosen according to the size of the feature map after convolution in the previous layer. If the feature map is larger than [3 × 3], C convolution kernels of size [3 × 3] are continued to be used, otherwise, C convolution kernels of the same size as the feature map are used. Thus, the 3-D EEG topographic map is convolved to produce an output of size [C × TP], the same size as input. After all convolutional layers, batch normalization is used to adjust the distribution of the data, and speed up the training progress.

\section{Implementation Details}
\begin{figure}[t]
\begin{center}
  \includegraphics[width=0.6\textwidth]{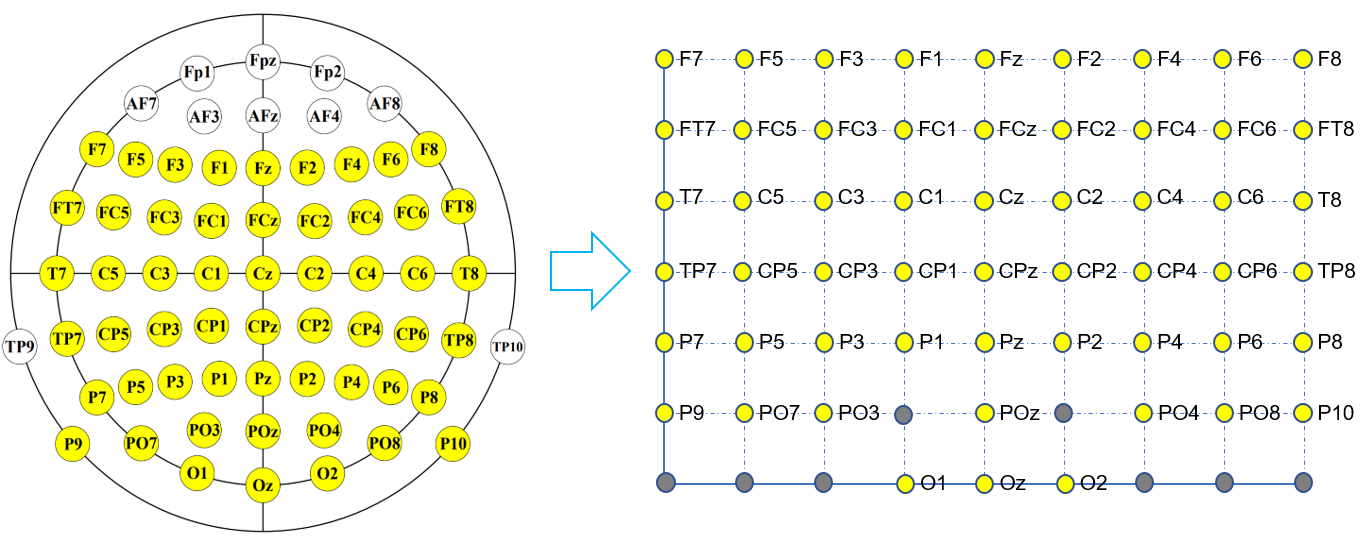}
  \caption{Correspondence between electrode locations and matrix coordinates on EBDSDD. The values at yellow points in the matrix are the potential values of their corresponding electrodes, and values of gray points are set to 0. A total of 55 electrodes are used, and the size of the corresponding matrix is [7 × 9].}
  \label{fig:Fig_2}
\end{center}
\end{figure}

\begin{figure}[t]
\begin{center}
  \includegraphics[width=0.6\textwidth]{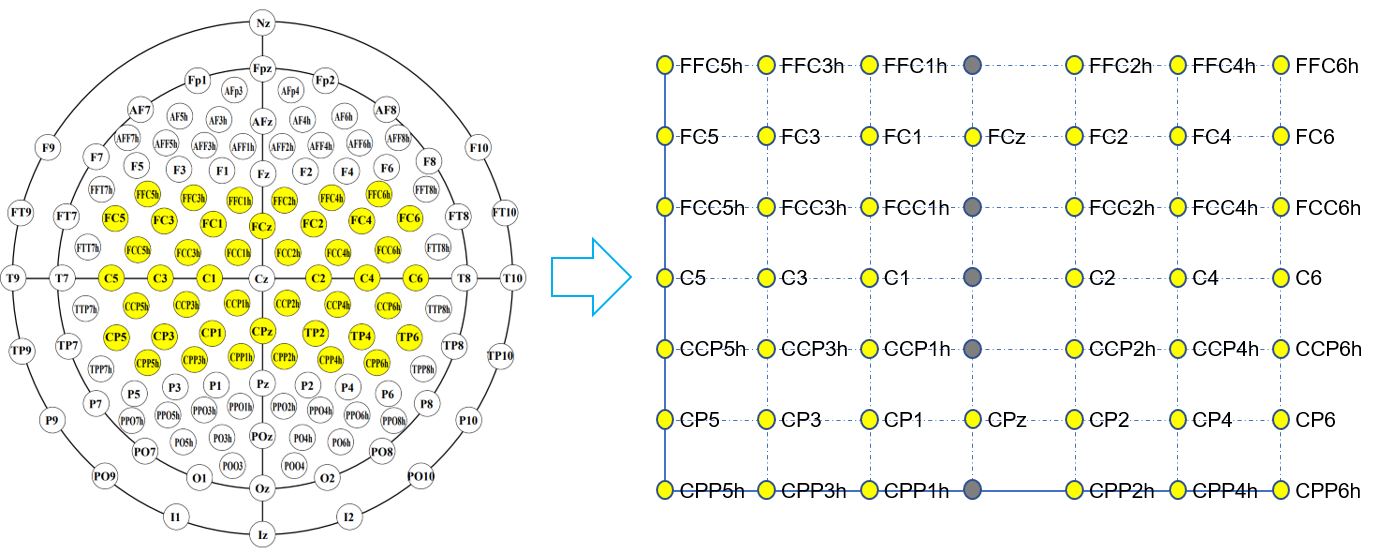}
  \caption{ Correspondence between electrode locations and matrix coordinates on HGD. The values at yellow points in the matrix are the potential values of their corresponding electrodes, and values of the gray points are set to 0. A total of 44 electrodes covering the motor cortex (all central electrodes except Cz, which is using as the recording reference electrode) are used, and the size of the corresponding matrix is [7 × 7].}
  \label{fig:Fig_3}
\end{center}
\end{figure}

Figures \ref{fig:Fig_2} and \ref{fig:Fig_3} the correspondence between electrode locations and 2-D matrix coordinates for EBDSDD and HGD, respectively. The values of the yellow points in the matrix are the potential values of the corresponding electrodes, and the values of the gray points are set to 0. Each matrix is a 2-D EEG topographic map for that moment. By arranging these 2-D matrices in the temporal order of the EEG signal, a 3-D EEG topographic map is formed. These two datasets are trained and tested for each subject using DeepConVNet, ShallowConvNet and EEGNet in their original form and with TRM-(5,5) or TRM-(3,3), respectively. For EBDSDD, we use a 4-fold cross-validation approach, with 50\% of each subject’s data as the training set, 25\% as the validation set, and the remaining 25\% as the test set. For HGD, the original data for each subject has been split into a training set and a test set. We keep the test set unchanged, and randomly select 80\% of the original training set as the new training set and the remaining 20\% as the validation set. Each subject's data is trained and tested 4 times, and the average is taken as the final result. For a fair comparison, we set the same training and testing conditions for all three CNNs and their varieties after using TRM. For each algorithm, the settings are the same except for the use or non-use of TRM. All algorithms are trained and tested on a computer with an NVIDIA 2080-Ti graphics card.  

Several aspects of the algorithms are set up as follows.

\begin{itemize}
    \item Adam optimizer is used with the weight-decay of 0.001, and the rest of the parameters are at default settings.
    \item Cross-entropy loss is used as a criterion.
    \item The batch size is set to 32. 
    \item The training epoch is set to 300 and a validation stopping strategy is used. The algorithm’s model with the lowest loss on the validation set is saved for testing.
    \item Both ShallowConvNet and EEGNet use the codes officially provided by braindecode for training and testing on both datasets (\cite{12}). On HGD, we use the code of DeepConvNet provided by the braindecode, while on EGDSDD, we adjust DeepConvNet with the setting recommended by (\cite{35}) since the size of the input data do not meet the minimum length requirement of the original form.
\end{itemize}

\section{Evaluation Metrics}
We comprehensively compare these algorithms using metrics such as classification accuracy, training loss, validation loss, the number of training epochs at lowest validation loss, and the time consumption analysis. Classification accuracy refers to the ratio of the number of correct classifications to the total number. Training loss and validation loss are the cross-entropy losses of the algorithms on the training and validation sets, respectively. We also count the number training epochs of each algorithm at the lowest validation loss with the aim of comparing which algorithm converges more easily. For the time consumption analysis, we use the time to train and validate 300 epochs with each algorithm.

\chapter{RESULTS}

\section{Classification Accuracy}
\textbf{Accuracy On EBDSDD.}
Table \ref{table:Table_1} shows the classification accuracies of 3 CNNs (ShallowConvNet, DeepConvNet and EEGNet) on EBDSDD with and without TRM, respectively. Results for each subject are mean value ± standard deviation of 4-fold cross-validation, and the p value in the table is calculated by two-tailed paired t-test. For DeepConvNet, the classification accuracies are improved for all 18 subjects after using TRM-(5, 5), with a maximum improvement of 22.6\% (Subject VPbba), and the average classification accuracy is improved by 6.54\% (p < 0.001); the classification accuracies of 17 subjects are improved after using TRM-(3, 3), with the highest increase of 27.40\% (Subject VPbba), only 1 subject has a slight decrease (Subject VPsaj with -0.46\%), and the average classification accuracy is improved by 7.76\% (p < 0.001). For EEGNET, after using TRM-(5, 5), the classification accuracies of 17 subjects are improved, with the highest increase of 2.92\% (Subject VPbad), only 1 subject shows a slight decrease in accuracy (Subject VPgaa with -0.42\%), and the average accuracy is improved by 1.72\% (p < 0.001); after using TRM-(3, 3), the classification accuracies of 15 subjects are improved, with the highest increase of 4.17\% (Subject VPbax), 3 subjects show an decrease in accuracy (Subject VPgaa with -0.85\%, Subject VPgab with -0.23, and Subject VPgal with -0.99\%), and the average accuracy is improved by 1.71\% (p < 0.001); For ShallowConvNet, after using TRM-(5, 5), the classification accuracies increase in all 18 subjects with a maximum of 5.82\% (Subject VPbba), and the average result is improved by 2.07\% (p < 0.001); after using TRM-(3, 3), the classification accuracies increase in 15 subjects with a maximum of 6.17\% (Subject VPbba), decreases in 2 subjects (Subject VPbax with -0.44\%, and Subject VPja with -0.72\%), remains unchanged in 1 subject (Subject VPgab), and the average result is improved by 2.17\% (p <0.001).

\begin{table}
\centering
\caption{CLASSIFICATION ACCURACIES OF DIFFERENT ALGORITHMS ON EBDSDD}
\label{table:Table_1}
  \resizebox{\textwidth}{!}{
\begin{tabular}{c c c c c c c c c c} 
\hline
\multirow{2}{*}{Subject} & \multicolumn{3}{c}{DeepConvNet} & \multicolumn{3}{c}{EEGNet}      & \multicolumn{3}{c}{ShallowConvNet}  \\ 
\cline{2-10}
                         & Original & TRM-(5,5) & TRM-(3,3) & Original & TRM-(5,5) & TRM-(3,3) & Original & TRM-(5,5) & TRM-(3,3)     \\ 
\hline
S1                       & 75.82    & 84.78     & 82.61     & 90.22    & 92.93     & 94.02     & 86.96    & 91.85     & 91.85         \\ 

S2                       & 88.51    & 95.27     & 96.17     & 95.05    & 97.97     & 97.97     & 93.02    & 97.75     & 98.65         \\ 

S3                       & 83.77    & 91.89     & 93.64     & 90.57    & 92.32     & 94.74     & 94.74    & 95.39     & 94.3          \\ 

S4                       & 55.82    & 78.42     & 83.22     & 88.7     & 91.44     & 91.44     & 84.93    & 90.75     & 91.1          \\ 

S5                       & 81.19    & 91.09     & 96.78     & 91.83    & 94.55     & 93.32     & 93.32    & 96.29     & 97.03         \\ 

S6                       & 94.7     & 97.67     & 98.09     & 98.94    & 98.52     & 98.09     & 97.25    & 98.52     & 98.09         \\ 

S7                       & 88.89    & 93.98     & 94.21     & 94.91    & 96.3      & 94.68     & 96.99    & 97.92     & 96.99         \\ 

S8                       & 91.52    & 95.09     & 95.54     & 96.88    & 97.77     & 97.77     & 97.32    & 97.99     & 97.99         \\ 

S9                       & 87.28    & 88.82     & 89.47     & 90.57    & 92.98     & 91.89     & 86.84    & 87.28     & 91.45         \\ 

S10                      & 89.95    & 96.81     & 98.04     & 95.83    & 98.04     & 98.04     & 95.83    & 97.06     & 97.79         \\ 

S11                      & 81.45    & 86.56     & 89.52     & 91.13    & 92.47     & 92.74     & 91.13    & 93.55     & 91.94         \\ 

S12                      & 80.69    & 84.9      & 89.6      & 93.81    & 94.06     & 92.82     & 94.31    & 94.8      & 95.3          \\ 

S13                      & 84.52    & 90.95     & 91.43     & 92.86    & 94.76     & 93.81     & 92.38    & 95.71     & 95.48         \\ 

S14                      & 81.37    & 89.86     & 91.04     & 93.63    & 95.05     & 96.7      & 95.05    & 95.52     & 96.23         \\ 

S15                      & 95.91    & 97.63     & 97.84     & 97.63    & 98.92     & 99.35     & 98.28    & 98.71     & 98.49         \\ 

S16                      & 89.56    & 91.99     & 93.45     & 93.2     & 95.87     & 95.87     & 95.87    & 96.6      & 95.15         \\ 

S17                      & 94.44    & 95.14     & 93.98     & 95.37    & 96.99     & 95.6      & 96.76    & 97.45     & 97.69         \\ 

S18                      & 70.43    & 82.69     & 80.77     & 89.9     & 91.11     & 93.03     & 86.78    & 91.83     & 91.35         \\ 
\hline
Average                  & 84.21    & 90.75     & 91.97     & 93.39    & 95.11     & 95.1      & 93.21    & 95.28     & 95.38         \\ 

p value                  & -----    & 4.54E-05  & 5.71E-05  & -----    & 3.90E-07  & 1.71E-4   & -----    & 2.43E-4   & 6.69E-4       \\
\hline
\end{tabular}
}
\end{table}

\textbf{Accuracy On HGD.}
Table \ref{table:Table_2} shows the classification accuracies of the 3 CNNs on HGD with and without TRM, respectively. For DeepConvNet, classification accuracies increase in 11 subjects after using TRM-(5, 5), with the highest increase of 22.33\% (Subject 9), and decreases in 3 subjects (Subject 2 with -0.94\%, Subject 11 with -2.66\%, and Subject 12 with -2.03\%), and the average classification accuracy increases by 6.05\% (p < 0.01); after using TRM-(3, 3), classification accuracies increase in 12 subjects with the highest increase of 22.49\% (Subject 9), decrease in 2 subjects (Subject 10 with -2.82\% and Subject 12 with -1.25\%), and the average result increases by 7.61\%. For EEGNet, 13 subjects show an increase in classification accuracy after using ERM-(5, 5), with the highest increase of 10.63\%, 1 subject shows a decrease (Subject 9 with -7.81\%), and the average classification accuracy is improved by 3.02\% (p < 0.05); all 14 subjects show increase in classification accuracy after using TRM-(3, 3), with the highest increase of 11.72\% (Subject 5), and the average classification accuracy is improved by 5.06\% (p < 0.001). For ShallowConvNet, after using TRM-(5, 5), the classification accuracies increase in 13 subjects with a maximum of 18.44\% (Subject 11), decreases in 1 subject (Subject 9 with -1.87\%), and the average classification accuracy is improved by 5.14\% (p < 0.01); after using TRM-(3, 3), the classification accuracies increase in 13 subjects with a maximum of 17.03\% (Subject 11), decreases in 1 subject (Subject 3 with -0.16\%), and the average classification accuracy is improved by 6.28\% (p < 0.001).

\begin{table}
\centering
\caption{CLASSIFICATION ACCURACIES OF DIFFERENT ALGORITHMS ON HGD}
\label{table:Table_2}
  \resizebox{\textwidth}{!}{
\begin{tabular}{c c c c c c c c c c} 
\hline
\multirow{2}{*}{Subject} & \multicolumn{3}{c}{DeepConvNet} & \multicolumn{3}{c}{EEGNet}      & \multicolumn{3}{c}{ShallowConvNet}  \\ 
\cline{2-10}
                         & Original & TRM-(5,5) & TRM-(3,3) & Original & TRM-(5,5) & TRM-(3,3) & Original & TRM-(5,5) & TRM-(3,3)     \\ 
\hline
S1                       & 57.81    & 62.81     & 63.91     & 54.84    & 58.75     & 56.72     & 67.34    & 72.66     & 77.81         \\ 

S2                       & 70.78    & 69.84     & 73.91     & 71.72    & 73.13     & 77.34     & 79.53    & 82.97     & 81.25         \\ 

S3                       & 83.13    & 90.16     & 92.03     & 92.81    & 95.31     & 96.72     & 95.16    & 95.31     & 95            \\ 

S4                       & 84.06    & 90        & 87.34     & 93.44    & 95.47     & 97.5      & 96.09    & 97.03     & 97.81         \\ 

S5                       & 62.97    & 67.66     & 82.5      & 70.16    & 72.97     & 81.88     & 81.25    & 90.16     & 89.06         \\ 

S6                       & 57.97    & 70.63     & 65.47     & 74.84    & 85.47     & 78.28     & 86.09    & 91.88     & 92.03         \\ 

S7                       & 59.69    & 67.66     & 68.28     & 65.31    & 69.53     & 71.41     & 75.16    & 81.09     & 84.38         \\ 

S8                       & 72.66    & 75.47     & 74.69     & 75.16    & 82.81     & 83.75     & 81.25    & 87.34     & 92.19         \\ 

S9                       & 47.51    & 69.84     & 70        & 76.56    & 68.75     & 84.84     & 78.75    & 76.88     & 85.63         \\ 

S10                      & 78.91    & 84.22     & 76.09     & 84.22    & 87.34     & 86.88     & 85.47    & 89.69     & 89.22         \\ 

S11                      & 62.66    & 60        & 68.28     & 73.44    & 75.66     & 79.38     & 78.28    & 96.72     & 95.31         \\ 

S12                      & 80.94    & 78.91     & 79.69     & 82.19    & 85.16     & 86.88     & 88.59    & 92.03     & 91.09         \\ 

S13                      & 59.38    & 65.63     & 67.81     & 75.78    & 80.94     & 77.81     & 79.06    & 87.66     & 85.47         \\ 

S14                      & 60.78    & 71.09     & 75.78     & 80.31    & 81.72     & 82.19     & 73.59    & 76.09     & 77.34         \\ 
\hline
Average                  & 67.09    & 73.14     & 74.7      & 76.48    & 79.5      & 81.54     & 81.83    & 86.97     & 88.11         \\ 

p value                  & -----    & 3.73E-3   & 1.75E-3   & -----    & 0.015     & 1.95E-05  & -----    & 1.72E-3   & 2.04E-4       \\
\hline
\end{tabular}
}
\end{table}

\section{Training Loss}
\begin{figure}[t]
\begin{center}
  \includegraphics[width=\textwidth]{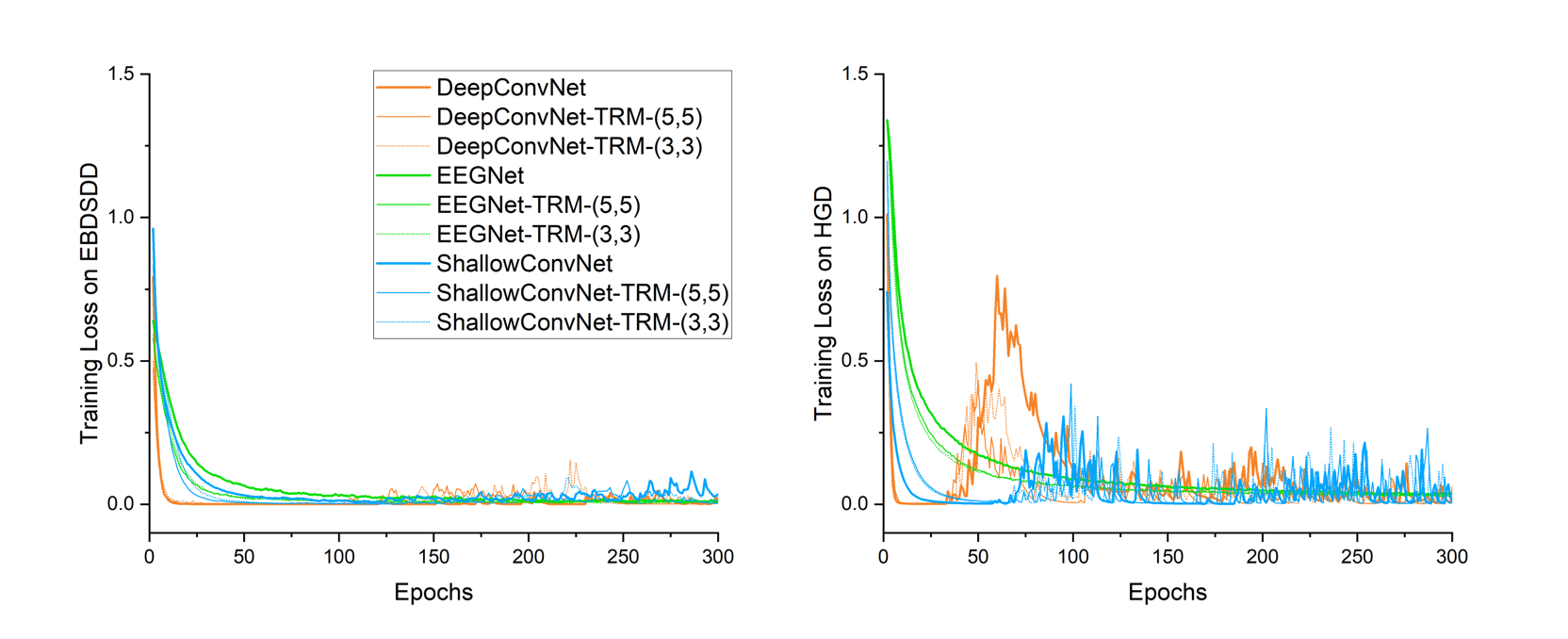}
  \caption{Training cross-entropy loss curves of different algorithms. The left and right panels show the average training loss curves for 18subjects on EBDSDD and 14 subjects on HGD, respectively. }
  \label{fig:Fig_4}
\end{center}
\end{figure}

Figure \ref{fig:Fig_4} shows the average training cross-entropy loss curves for each algorithm. With TRM, the training loss curves of all three CNNs follow roughly the same trend as their do in their original forms on both datasets. DeepConvNet and its varieties have the fastest decline in the training loss curve, followed by the ShallowConvNet, while EEGNet and its varieties’ curves are relatively flat. On EBDSDD, the losses of all algorithms are relatively small after a period of training, which is related to the fact that this dataset is a 2-classification task with a high classification accuracy. We find that the training loss curves are not smooth, especially DeepConvNet, ShallowConvNet and their varieties on HGD, which is related to the use of weight-decay. By imposing certain restrictions on the learning weights, the overfitting problem can be mitigated to some extent. The training loss curves of EEGNet and its varieties show a relatively smooth downward trend on both datasets, which we believe is related to the fact that it has fewer trainable parameters compared to the other two algorithms. 

\section{Validation Loss}
\begin{figure}[t]
\begin{center}
  \includegraphics[width=\textwidth]{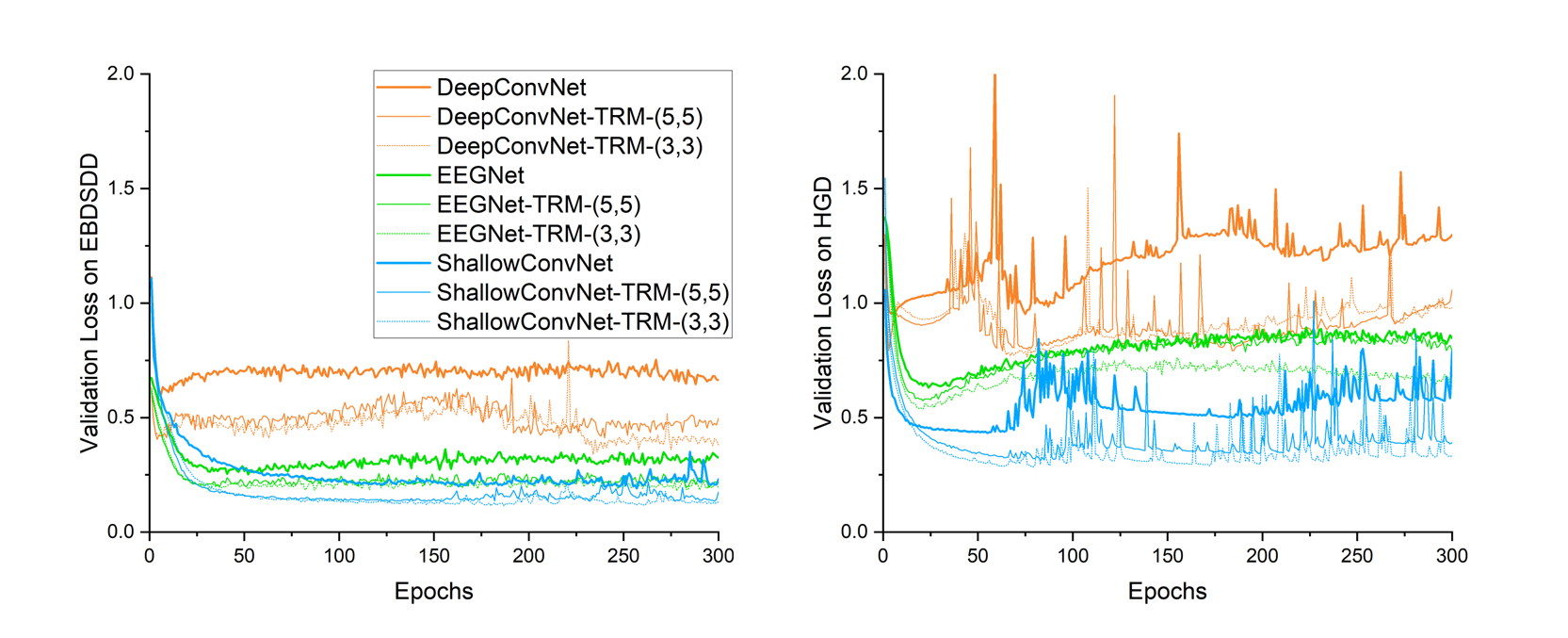}
  \caption{Validation cross-entropy loss curves of different algorithms. The left and right panels show the average validation loss curves for 18 subjects on EBDSDD and 14 subjects on HGD, respectively.}
  \label{fig:Fig_5}
\end{center}
\end{figure}

Figure \ref{fig:Fig_5} shows the average validation loss curves for each algorithm. Compared with the training loss curve, the validation curve provides a more accurate reflection of the classification and convergence performance. Similar to what we see on the training loss curve, either with TRM-(5, 5) or TRM-(3, 3), the validation loss curves of all three CNNs show a similar trend to their original forms on both datasets. After the first few epochs of training, the validation loss of DeepConvNet is significantly higher than that of the other two algorithms on both datasets, and even with TRM, its validation loss is still higher than the other two methods, a result consistent with the classification accuracy. On EBDSDD, EEGNet, EEGNet-TRMs, ShallowConvNet and ShallowConvNet-TRMs all have good validation loss curves, while on the HGD, ShallowConvNet and ShallowConvNet-TRMs are better than EEGNet and EEGNet-TRMs. With TRM, the validation loss curves of all 3 algorithms show a decreasing trend, indicating that the TRM is able to improve their classification performance. 

\section{Number of Training Epochs at The Lowest Validation Loss}
\begin{figure}
\begin{center}
  \includegraphics[width=\textwidth]{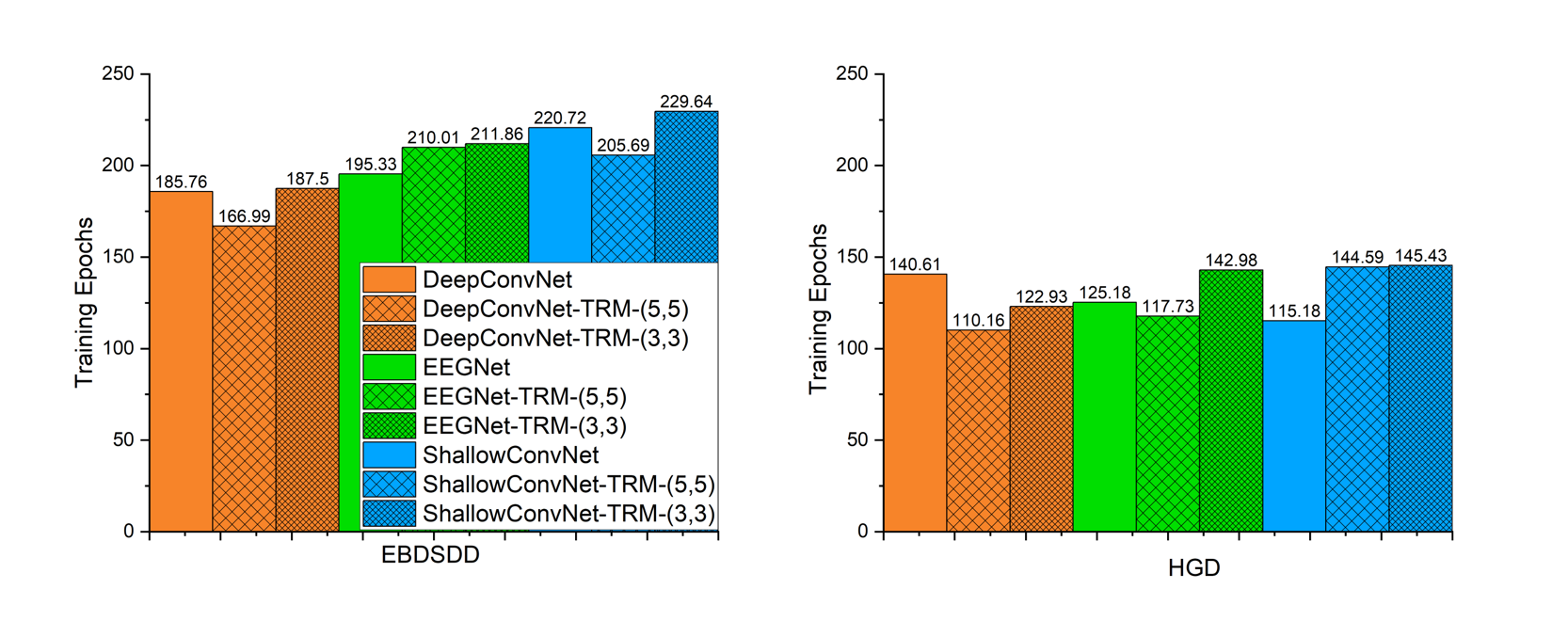}
  \caption{Average number of training epochs at the lowest validation loss for each algorithm. The left and right panels show the average number of training epochs at the lowest validation loss for 18 subjects on EBDSDD and 14 subjects on HGD, respectively. }
  \label{fig:Fig_6}
\end{center}
\end{figure}

Figure \ref{fig:Fig_6} the average number of training epochs at the lowest validation loss for each algorithm. The average number on EBDSDD is overall higher than that on HGD. With TRM-(5, 5), the average number of training epochs at the lowest validation loss decreases for DeepConvNet (p=0.25) and ShallowConvNet (p=0.22) and increases for EEGNet (p=0.09) on EBDSDD; on HGD, the average number decreases for DeepConvNet (p=0.07) and EEGNet (p=0.72) and increases for ShallowConvNet (p=0.06). With TRM-(3, 3), the average number of training epochs at the lowest validation increases for DeepConvNet (p=0.87), EEGNet (p=0.13) and ShallowConvNet (p=0.46); on HGD, the average number decreases for DeepConvNet (p=0.29), and increases for EEGNet (p=0.23) and ShallowConvNet (p=0.02). The p value is calculated by two-tailed paired t-test.

\section{Time Consumption Analysis}
\begin{figure}
\begin{center}
  \includegraphics[width=\textwidth]{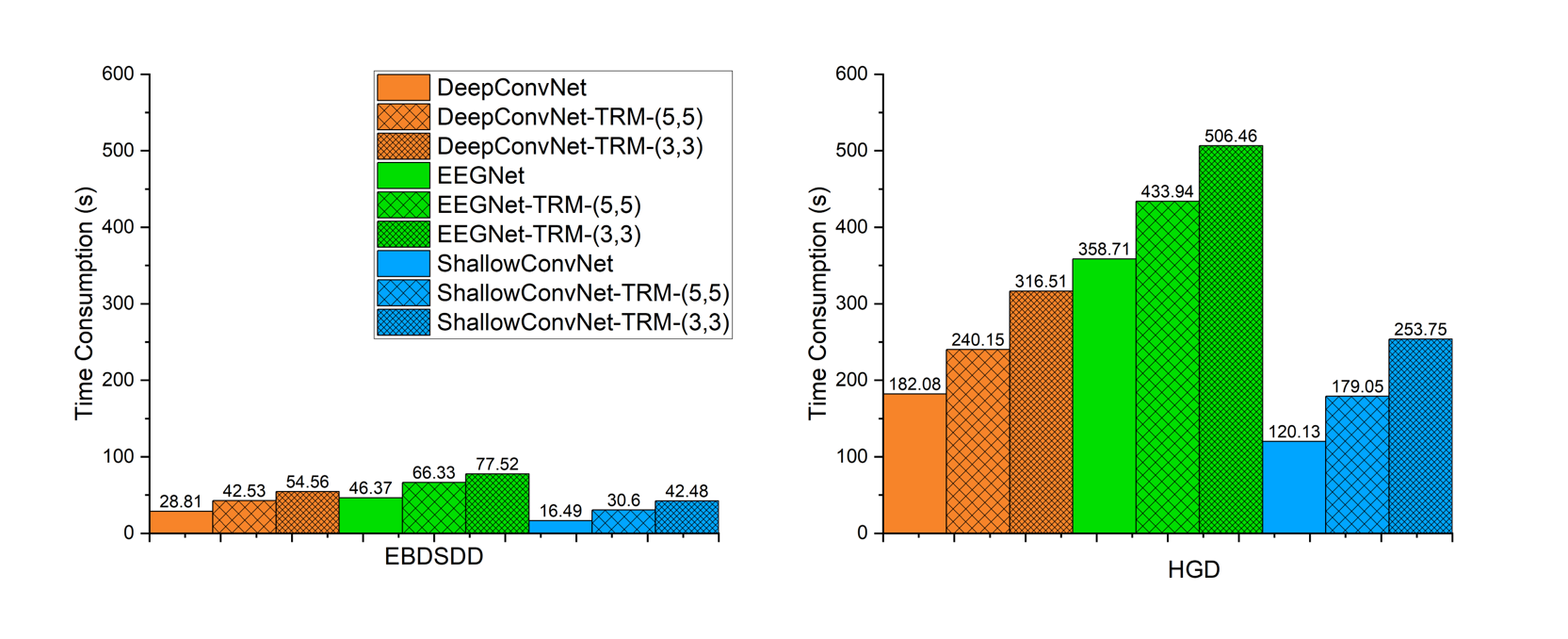}
  \caption{Average time consumption for training and validation 300 epochs with each algorithm. The left and right panels show the average time consumption for18 subjects on EBDSDD and 14 subjects on HGD, respectively.}
  \label{fig:Fig_7}
\end{center}
\end{figure}

Considering the practicality in BCIs, we need to consider not only the classification accuracy, but also the execution time of the algorithm. Due to the addition of TRM, the execution time of the algorithm definitely increases. Figure \ref{fig:Fig_7} shows the average time consumption of the algorithm for 300 training and validation epochs on each subject. On both datasets, ShallowConvNet consumes less time than DeepConvNet, and DeepConvNet consumes less time than EEGNet. With TRM-(5, 5), the time for training and validation of ShallowConvNet, DeepConvNet and EEGNet 300 epochs increases by about 86\%, 48\%, and 43\% on EBDSDD, and about 49\%, 32\%, and 21\% on HGD, respectively. With TRM-(3, 3), the time of ShallowConvNet, DeepConvNet and EEGNet increases by about 158\%, 89\%, and 67\% on EBDSDD, and about 111\%, 74\%, and 41\% on HGD, respectively. The execution time of ShallowConvNet-TRM-(5, 5) is comparable to DeepConvNet, while DeepConvNet-TRM (5, 5) is less than EEGNet. ShallowConvNet-TRM-(3, 3) has less execution time than EEGNet on both datasets, and DeepConvNet-TRM-(3, 3) has more than EEGNet on EBDSDD and less than EEGNet on HGD. The increased time consumption is relatively acceptable.

\chapter{Discussion}
The raw EEG signal is generally represented as a 2-D matrix form of [C (Channel) × TP (Time Points)]. In a CNN with the raw EEG signal as input, if there is no representation module about the electrode spatial topology in the network, the EEG signal will be processed as a tensor similar to a 2-D picture, and the spatial topological information of the electrode will be ignored. The topographic map is a representation of the EEG signal as a 2-D or 3-D image, depending on the spatial topology of the electrodes (their locations on the scalp) (\cite{27}). Topographic maps can be constructed using either the raw EEG signal (\cite{23, 48, 49}), or extracted features (\cite{13, 46, 47, 50, 51}).  If a CNN use the EEG topographic map constructed from extracted features as input, its performance depends on the quality of the features, which often requires substantial expertise and a priori knowledge. In studies of CNNs using topographic maps constructed from raw EEG signals, to the best of our current knowledge, all the networks have been designed according to the needs of the task. Given that there are already many excellent CNNs that use the raw EEG signal as input, it is possible to achieve the purpose of using spatial topological information of the EEG by simply adding a module without changing the structure of the original network.

Therefore, we designed the EEG Topographic Representation Module (TRM). By mapping the raw EEG signal into a 3-D topographic map, we make the input contain the spatial topological information of the electrodes. For those points without corresponding electrodes, we adopted the practice of (\cite{48}) and directly set them to 0 instead of using interpolation, because in our experiments, we find that interpolation does not lead to performance improvement but increase time consumption. We perform a convolution operation on the 3-D EEG topographic map to transform it into an output of the same size and dimensions as the input. Depending on the size of the convolutional kernels used, two convolutional strategies, TRM-(5,5) and TRM-(3,3), are chosen to analyze the impact of the size of the convolutional kernels on the classification performance of the algorithm. Such a design takes advantage of the powerful feature learning capabilities of deep learning while using a variety of existing excellent CNNs in the EEG-based BCI.

We choose three widely used CNNs, namely ShallowConvNet with a shallow structure, DeepConvNet with a deep structure, and the compact EEGNet. And for datasets, we use two datasets from different types of BCI paradigms. It is hoped that these practices make our results representative. We are very pleased to find that each CNN with TRM had a strong similarity to the original network, both in terms of the training loss curve and the validation loss curve, indicating that the properties of the original CNN are largely preserved with the use of TRM. Under the same training and test conditions, the validation loss curves of all three CNNs show a downward trend after using TRM-(5, 5) or TRM-(3, 3) on both datasets, which to some extent indicates the ability of TRM to improve the classification performance of EEG-based CNNs.

\begin{figure}
\begin{center}
  \includegraphics[width=\textwidth]{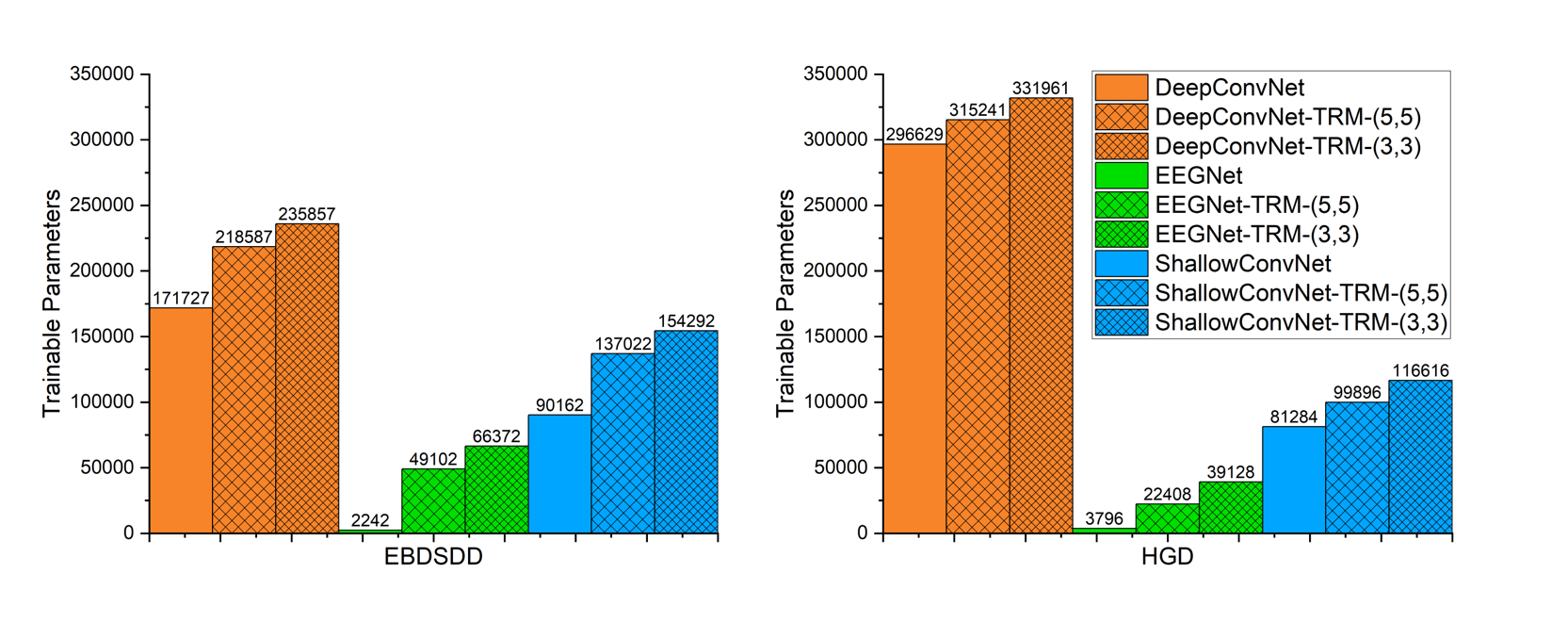}
  \caption{The number of trainable parameters for each algorithm. The left and right panels show the number of trainable parameters of the algorithm when used on EBDSDD and HGD, respectively.}
  \label{fig:Fig_8}
\end{center}
\end{figure}

As the size of the 3-D topographic map obtained from the raw EEG signal and the size of the convolutional kernel used vary, the number of trainable parameters of the TRM changes accordingly. Figure \ref{fig:Fig_8} shows the number of trainable parameters of the 3 CNNs with or without TRM. In this paper, the number of trainable parameters for TRM-(5, 5) is 46860 when used on EBDSDD and 18612 on HGD, and TRM-(3, 3) is 64130 when used on EBDSDD and 35332 when used on HGD.

ShallowConvNet has a simple structure, consisting of only a temporal convolutional layer, a spatial convolutional layer, a pooling layer and a dense layer, and the number of parameters is relatively moderate, so it performs the fastest among all three CNNs. Even with TRM-(5, 5), the time it consumes is still comparable to DeepConvNet and significantly less than EEGNet. When with TRM-(3, 3), it consumes more time than DeepConvNet, but still less than EEGNet. On EBDSDD, ShallowConvNet and its varieties achieve the best results on 10 out 18 subjects, with ShallowConvNet-TRM-(5, 5) accounting for 6 and ShallowConvNet-TRM-(3, 3) for 4. The average classification accuracy of ShallowConvNet is lower than that of EEGNet, while ShallowConvNet-TRM-(5, 5) and ShallowConvNet-TRM-(3, 3) are higher than that of EEGNet. Considering the classification accuracy and the execution time of the algorithm, we recommend ShallowConvNet-TRM-(5, 5). On HGD, ShallowConvNet and its varieties achieve the highest classification accuracy on 12 out of 14 subjects, with ShallowConvNet-TRM-(5, 5) and ShallowConvNet-TRM-(3, 3) each accounting for 6 of them, and the ShallowConvNet-TRM-(3, 3) achieve the highest average classification accuracy. Inspired by FBCSP, ShallowConvNet often has a better classification performance in spontaneous EEG decoding (\cite{11,35}).  From the validation loss curve on HGD, we can also find that ShallowConvNet, ShallowConvNet-TRM-(5, 5) and ShallowConvNet-TRM-(3, 3) have a smaller validation loss than other methods. Thus, ShallowConvNet-TRM-(3, 3) is the recommended algorithm to be used on HGD.

DeepConvNet has a relatively deep structure and a large number of trainable parameters, which often requires a large number of samples and uses certain skills to train. With fewer samples, it is often prone to overfitting (\cite{11,18,35}). The classification results of DeepConvNet are poor on both datasets, and the validation loss curves also indicates that it has a larger validation loss compared to other methods. Although the classification accuracy of DeepConvNet is improved after using TRM, it is still lower than that of ShallowConvNet and EEGNet, which may be related to our smaller sample size. The time consumption of DeepConvNet is higher than that of ShallowConvNet. With TRM-(5, 5), its time consumption increases, but is still lower compared to EEGNet. When TRM-(3, 3) is used, its time consumption exceeds that of EEGNet on EBDSDD and remains lower than that of EEGNet on HGD. ShallowConvNet and its varieties are not recommended when the trainable samples are small.

By using depth-wise and separable convolutions, and omitting the dense layer, the number of trainable parameters in EEGNet is at least one order of magnitude smaller than that in the other two algorithms, which greatly alleviates the overfitting problem that often occurs in deep learning (\cite{11}). Compared to the other two methods, EEGNet has a smoother training loss curve and validation loss curve, and the same is true when TRM is used. Although EEGNet has the least number of trainable parameters, it has a highest time consumption among the 3 CNNs. With TRM, its trainable parameters are significantly improved. On EBDSDD, EEGNet and its use of TRM achieve the best results in 8 out of 18 subjects, with EEGNet, EEGNet-TRM-(5, 5), and EEGNet-TRM-(3, 3) each accounting for 1, 2, and 5 of them. Compared with EEGNet, the average classification accuracy of EEGNet-TRM-(5, 5) is improved by 1.72\% (p-value < 0.001), which is very valuable in such a high classification accuracy. On HGD, EEGNet and its varieties achieve the best results of 2 out of 14 subjects, and all of them are achieved by EEGNet-TRM-(3, 3). The average classification accuracy of EEGNet with TRM-(3, 3) is improved by 5.06\% (p < 0.001), but it is still lower than that of ShallowConvNet.

By the analysis of the TRM structure, we find that a large portion of the increased time is spent on mapping the raw EEG signal into 3-D topographic map. If the EEG signal is collected as a 3-D matrix (3-D topographic map), TRM can omit the mapping process. In this case, the additional time consumption is also reduced.

\chapter{Conclusion}
In this paper, we introduce an EEG Topographic Representation Module (TRM). This module consists of a mapping block and a convolution block, and has an output the same size as its input.  Any CNN with the raw EEG signal as input can use TRM. We selected 3 representative CNNs and tested them on 2 different types of BCI datasets. Results show that the classification accuracy of all 3 algorithms is improved after using TRM. Next, we intend to train and test TRM on more CNNs and more datasets for a further validation. In addition, we want to improve the TRM structure to reduce its time consumption.

\newpage

\printbibliography
\end{document}